\documentclass{aa}
\usepackage{graphicx}
\usepackage{txfonts}
\begin{document}

   \title{Catching VY Sculptoris in a low state 
\thanks{Based on observations collected at the European Southern Observatory, Paranal (programmes 282.D-5017 and 081.D-0318). The reduced spectra and radial velocities are available in electronic form at the CDS via anonymous ftp to cdsarc.u-strasbg.fr (130.79.128.5)
or via http://cdsweb.u-strasbg.fr/cgi-bin/qcat?J/A+A/ }}
   \subtitle{}
   \author{
L.\ Schmidtobreick \inst{1} \and
E.\ Mason \inst{2} \and
S.\ B.\ Howell \inst{3} \and
K.\ S.\ Long \inst{4} \and
A.\ F.\ Pala \inst{5} \and
S.\ Points \inst{6} \and
F.\ M.\ Walter \inst{7}}
   \institute{
European Southern Observatory, Casilla 19001, Santiago 19, CL\\
\email{lschmidt@eso.org} \and
INAF-OATS, Via G.B. Tiepolo 11, 34143, Trieste, IT \and
NASA Ames Research Center, PO Box 1, M/S 244-30, Moffett Field, CA 94035, USA\and
Space Telescope Science Institute, 3700 San Martin Drive, Baltimore, MD 21218, USA\and
Department of Physics, University of Warwick, Coventry CV4 7AL, UK\and
Cerro Tololo Inter-American Observatory, Casilla 603, La Serena, CL\and
Department of Physics and Astronomy, SUNY Stony Brook, Stony Brook, NY 11794-3800, USA}
   \date{xxx}
\abstract
{In the context of a large campaign to determine the system parameters of high mass transfer cataclysmic variables, we found VY\,Scl in a low state in 2008.}
{Making use of this low state, we study the stellar components of the binary with little influence of the normally dominating accretion disc.}
{Time-resolved spectroscopy and photometry of VY\,Scl taken during the low state
are presented. We analysed the light-curve and radial velocity curve and use  time-resolved spectroscopy to calculate Doppler maps of the dominant emission lines.}
{The spectra show narrow emission
lines of H$\alpha$, H$\beta$, He\,{\sc i}, Na\,{\sc i}\,D, and Fe\,{\sc ii,} as well as faint TiO
absorption bands that trace the motion of the irradiated secondary star,
and H$\alpha$ and He{\sc i} emission line wings that trace the motion of the white dwarf. From
these radial velocities, we find an
orbital period of 3.84\,h, and put constraints on binary parameters such as
the mass ratio $M_2/M_1$ of 0.43 and the inclination of $15^\circ$. With a secondary's mass between 0.3 and 0.35 M$_\odot$, we derive the mass for the white dwarf as $M_1 = 0.6-1.1$\,M$_\odot$. }{}
   \maketitle
%

\section{Introduction}
VY Sculptoris is a cataclysmic variable (CV), an interacting binary comprising a white dwarf primary and a Roche-lobe filling red dwarf secondary star \cite[for more information on CVs, see][]{warner95-1}. It belongs to the class of nova-like stars which
are weakly or non-magnetic CVs with a high mass transfer rate  feeding
a hot, steady-state accretion disc. The disc dominates the light emission of the binary and prevents the observation of the two stellar components.
However, many of these nova-like stars show
occasional low states in which the mass transfer is diminished or even
completely suppressed \citep[see][for a review and more references]{schmidtobreick17-01}. This leads to a weakness or absence of the 
usually dominating accretion disc flux and produces a drop in brightness of 
3--6\,mag in the optical. 
Thus, low states provide
a unique opportunity to study the white dwarf and/or the donor star in
such systems; to determine system parameters such as temperature,
mass, radius, and stellar types from the spectrum;
and  by following their radial velocity curves over time 
to derive the dynamical masses of these components  \citep[see e.g.][]{rodriguez-giletal15-1}.

VY\,Scl is the prototype of this subgroup of stars: 
typically in bright state 
at V magnitude of $\sim$13--14, it has occasionally been reported to be as 
faint as V$\sim$18 \citep{warner+vancitter74-1}.
VY Scl type stars are mostly found with 
orbital periods between 3\,h and 4\,h
\citep{king+cannizzo98-1, warner99-1}.  Warner asserts that
all nova-like stars in this period range which have been regularly
monitored show VY\,Scl low states.

The reason why VY\,Scl stars enter a low mass transfer state is not well understood. Most explanations  to subdue or stop the mass transfer assume magnetic cycles in the secondary star and connect the mass transfer to the  stellar activity.
Originally, the presence of star spots on the L1 point were suggested by
\cite{livio+pringle94-1} as the subduing mechanism, while later explanations
propose a decrease in the secondary star's size due to magnetic activity
\citep{howell04-1}.

While in decline, the previously hot and steady-state accretion disc 
of the nova-like star will lose mass onto the white dwarf and become fainter 
and cooler until it disappears.
\cite{leachetal99-1}, \cite{hameury+lasota02-1}, and \cite{hameury+lasota05-2} discuss that to prevent disc instability outbursts during this decline phase, the disc has to be truncated either through irradiation by a hot white dwarf or via an intermediate magnetic field on the white dwarf.

VY\,Scl itself was discovered as a variable blue star by \cite{luyten+haro59-1}.
The first spectra were presented by \cite{burrell+mould73-1}.
They report hydrogen and helium lines in emission, but no trace of
any absorption lines or other indications of a late-type secondary.
In 1983, VY\,Scl went into a low state  (November 1983, V$\sim$16 mag)
 and was observed spectroscopically
by \cite{hutchings+cowley84-1} who estimated the orbital period 
$P=3.984$\,h. 
This value was later challenged by \cite{martinez-paisetal00-1} who
 observed the object in bright state (V$\sim$13--14 mag) and
give a most likely period of 5.59\,h. They also claim a possible
triple nature for the system.

In the framework of a large monitoring project to observe VY\,Scl
stars in low state with the aim of getting dynamical masses of these objects,
VY\,Scl was caught in a low state (V$\sim$18 mag)
 in October--November 2008 and was observed in 
time-resolved spectroscopy mode. During the observing
epochs and in particular during the time-resolved spectroscopy runs,
mass accretion and a partial accretion disc were present to various degrees and continued to veil the two stars. However, we did obtain sufficiently red spectra which revealed a TiO bandhead from the secondary star. Analysing the variation in the
emission and absorption lines allowed us to determine the orbital period and to constrain
other parameters such as the  mass ratio and the binary inclination.

\section{ Light curve}
\label{lightcurve}
\begin{figure}
\resizebox{9cm}{!}{\includegraphics{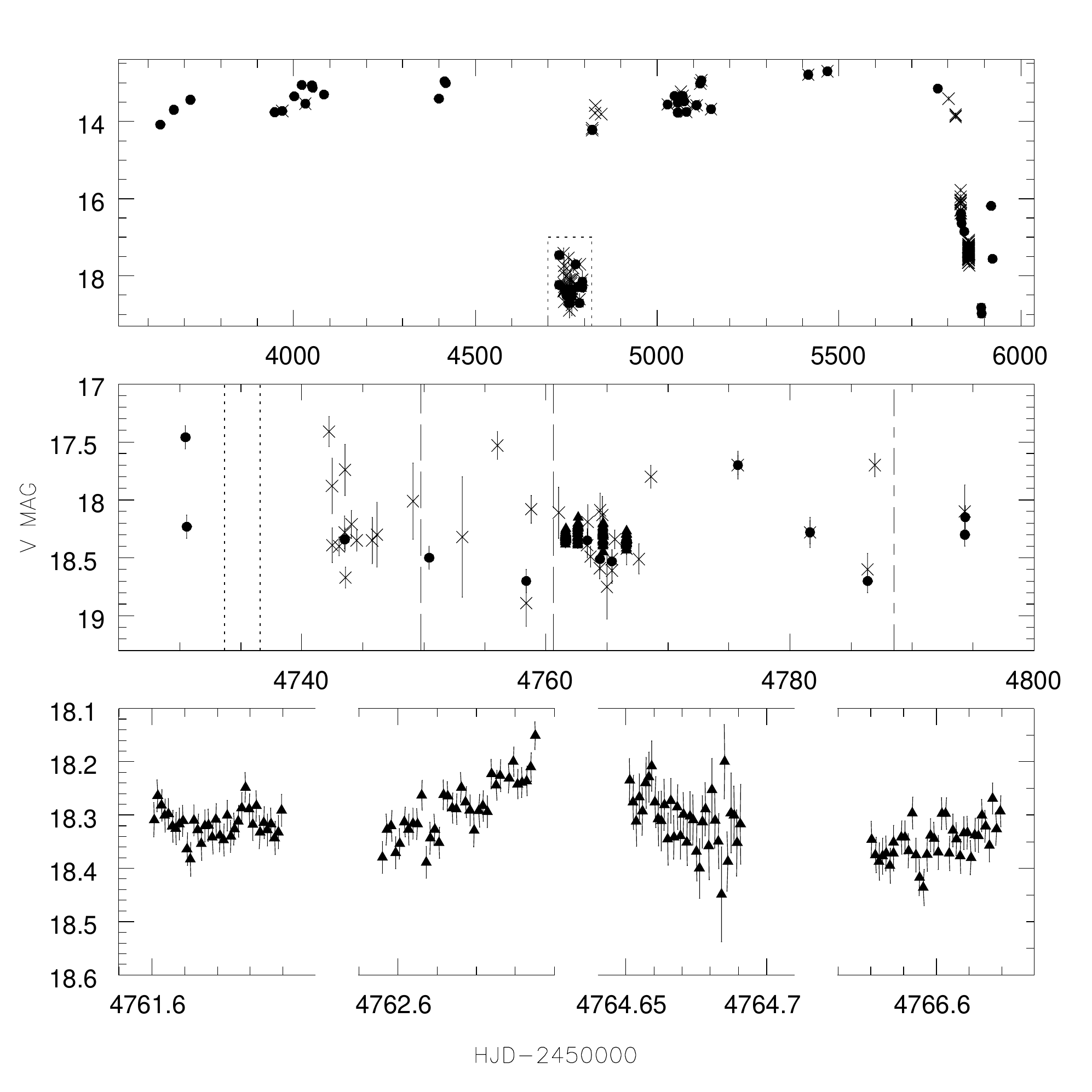}}
\caption{\label{vyscl_lc} Top: Long-term $V$-band light curve of VY\,Scl showing 4.17\,years of photometric coverage (01 September 2006 to 31 October 2010). The `x' symbols are observations from the AAVSO data base, while black circles are from the M1 Group (Spanish amateur astronomers).  Middle: Enlarged view of the data points in the dotted box marked in the top panel. The `x' symbols and black circles  are as above. Black triangles refer to the SMARTS V band photometry (see Table \ref{obstab}). The vertical dotted lines mark the start of the two FORS2 runs, the dashed lines mark the start of the two Goodman runs, and the long dash--short dashed line mark the start of the FORS1 run. Bottom: Zoomed in view of the SMARTS photometry.}
\end{figure}

VY\,Scl is one of the nova-like stars that are regularly monitored for
low states.
In Fig.\ \ref{vyscl_lc} the long-term $V$-band light curve covering
4.17\,years
from 01 September 2006 to 31 October 2010 is plotted.
All data were taken by amateur astronomers either from AAVSO (slanted crosses in Fig.\ref{vyscl_lc}, top panel) or from the Spanish M1 Group (solid black circle in Fig.\ref{vyscl_lc}, top panel).
In September 2008, the brightness of VY\,Scl decreased by about 4.5\,mag from an average of $V=13.33$ to $V=18.2$.
VY\,Scl stayed in this low state for about 70\,days before going back to its normal brightness at the end of November 2008.
During this low state, time-resolved photometric observations were taken with the ANDICAM dual-channel imager on the SMARTS/CTIO
1.3m telescope on four nights between 21 and 26 October 2008
(see Table\,\ref{obstab} and Fig.\ref{vyscl_lc}, bottom panel, for details).
No source was detected in the infrared channel which had a
limiting detection magnitude near K$\sim$16. The general data reduction
procedures are outlined in \cite{walter+12-1}.
Differential $V$-photometry was performed
against the seven brightest stars in the $\approx$6~arcmin
field of view. Only data with uncertainties $<0.05$~mag
were considered in the further analysis.

Similarly to other VY\,Scl stars, VY Scl
showed aperiodic brightness variations during its low state.
In the past, such
variability was interpreted as sporadic mass transfer events or
stunted dwarf nova outbursts \citep[and references therein]{rodriguez-giletal12-1}.
However, our sampling is too sparse to draw any such conclusions for VY\,Scl.

We used the four sets of SMARTS data that were taken during the low state to search for a
periodic signal using the Scargle \citep{scargle81-1} and
analysis-of-variance (AOV) \citep{schwarzenberg-czerny89-1} algorithms
as implemented in MIDAS. Both power spectra show many maxima of nearly equal
power between 6\,h and 2.5\,h. The highest peak, though not statistically significant, gives a period of 3.85\,h. 
Using instead the shortest string
method \citep{dworetsky83-1}, we find a period near 3.70 hours;  this peak is also not statistically significant. The lack of a significant periodic signal in the SMARTS
photometric data suggests either values for the orbital period that are significantly longer than the 1\,h observing windows or a
rather low inclination of the system.

\begin{table*}
\small
  \caption{\label{obstab} Log of observations.}
  \begin{tabular}{lllllllllll}
  \hline
 UT date & UT start & UT end & exptime & \# exp. & instrument & set-up & $\lambda$-range [\AA]& R[\AA] & sky & seeing \\
\hline
2008-09-24 & 3:33 & 3:55 & 600\,s & 1 & FORS2 & 600B/1.0'' & 3300-6210 & 6.0 & CLR & 0.9" \\
2008-09-24 & 4:00 & 4:16 & 600\,s & 1 & FORS2 & 1200R+GG435/0.7'' & 5750-7250 & 2.2 & CLR & 0.8" \\
2008-09-27 & 2:21 & 2:41 & 600\,s & 2  & FORS2 & 1200R+GG435/0.7'' & 5750-7250 & 2.2 & CLR/PHO & 0.7" \\
2008-09-27 & 2:41 & 8:42 & 300\,s & 60 & FORS2 & 1200R+GG435/0.7'' & 5750-7250 & 2.2 & CLR/PHO & 0.8"-1.3" \\
2008-10-10 & 5:20 & 6:25  & 1200\,s & 3 & Goodman & RALC300/1.03" & 4550-8920& 9.2 & CLR/PHO & 0.8"\\
2008-10-21 & 4:38 & 4:58  & 1200\,s & 1 & Goodman & KOSI600B/1.03" & 3500-6160& 4.4 & CLR/PHO & 0.8"\\

2008-10-22 & 2:25 & 3:21 & 48\,s & 36 & ANDICAM & V,K & & & & \\
2008-10-23 & 2:18 & 3:15 & 48\,s & 35 & ANDICAM & V,H & & & & \\
2008-10-25 & 3:38 & 4:35 & 48\,s & 36 & ANDICAM & V,H & & & & \\
2008-10-27 & 1:55 & 2:53 & 48\,s & 35 & ANDICAM & V,H & & & & \\
2008-11-18 & 0:13 & 4:06 & 600\,s & 21 & FORS1 & 600V+GG375/1.0'' & 4500-7500 & 5.9 & CLR/PHO &  0.7"\\
\hline
\end{tabular}
\end{table*}

\section{Spectroscopic observations and data reduction}

Time series spectroscopy of VY\,Scl was performed on 27 September 2008 at UT1+FORS2
and on
18 November 2008 at UT2+FORS1 of the VLT on Cerro Paranal (see \citealt{appenzelleretal98-2} for details on the FORS instruments).
During the first run VY\, Scl was followed for about 6.5 consecutive
hours taking exposures of 5 minutes each, while on the second run
VY\,Scl was followed for nearly 4\,h with exposures of 10 minutes each.
Two snapshots were taken with UT1+FORS2 in the blue and in the red optical wavelengths a few days before the FORS2 time series of September 2008. Two more snapshots were taken on 10 and 21 October 2008 at the 4.1m SOAR telescope using the Goodman spectrograph \citep{clemens+04-1}.
All observing runs were executed during the same
low state.
A detailed log of the observations is given in Table \ref{obstab}.

All data were reduced following standard procedures using {\sc iraf}\footnote{{\sc iraf} is distributed by
the National Optical Astronomy Observatories} or {\sc molly}\footnote{Tom Marsh's {\sc molly} package is available at http://deneb.astro.\\warwick.ac.uk/phsaap/software/}. 
The September data were reduced using both packages,
while the October and November data set was reduced using only {\sc iraf} and its
tasks for long-slit spectroscopy. All spectra were corrected for bias
and flats, optimal extracted \citep{horne86-1}, and wavelength and flux calibrated. Flux calibration was not applied to the time series on 27 September since no spectrophotometric standard was taken on the same night. 
The November spectra were taken in photometric conditions and, using a spectrophotometric standard star, we were able to flux calibrate them.
During the analysis of the November data sets we discovered that the grism 600V of FORS1 shows second-order overlap beyond 7000A. Therefore, we ignored the red part of the November spectra in our analysis. 

\begin{figure}
\resizebox{8.9cm}{!}{\includegraphics{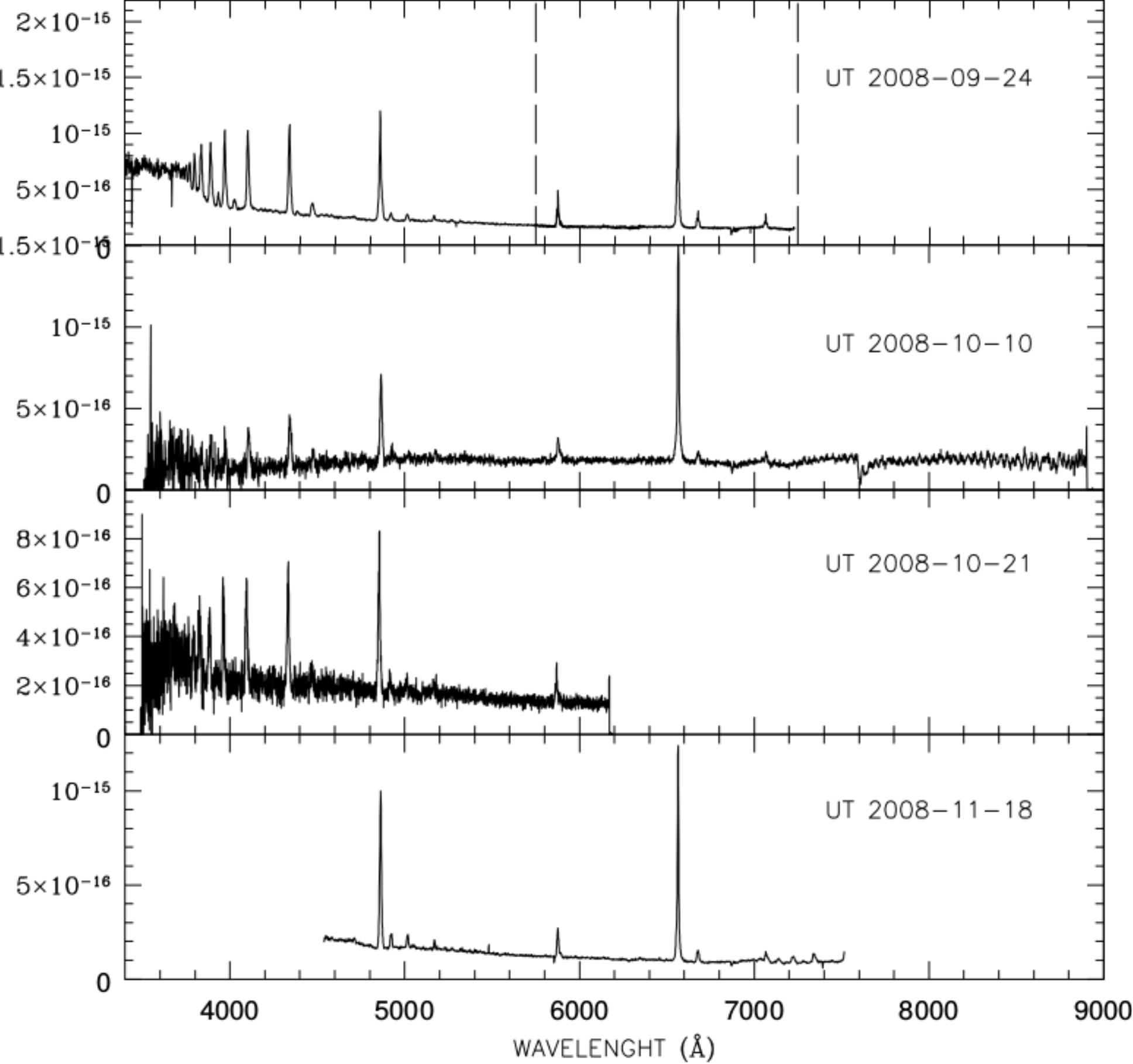}}
\caption{\label{flxedSPC} Sample  fluxed spectra (ergs/cm$^2$/s/\AA)  from the spectroscopic observations in Table 1. The UT date of each observing run is indicated in the figure itself. The dashed vertical lines in the top panel delimit the wavelength range of the time series from 27 October.  The emission lines in the November 2008 spectrum (bottom panel) redward of 7000\AA\, are due to second-order overlap (see text).}
\end{figure}

\begin{figure}
\resizebox{9.0cm}{!}{\includegraphics{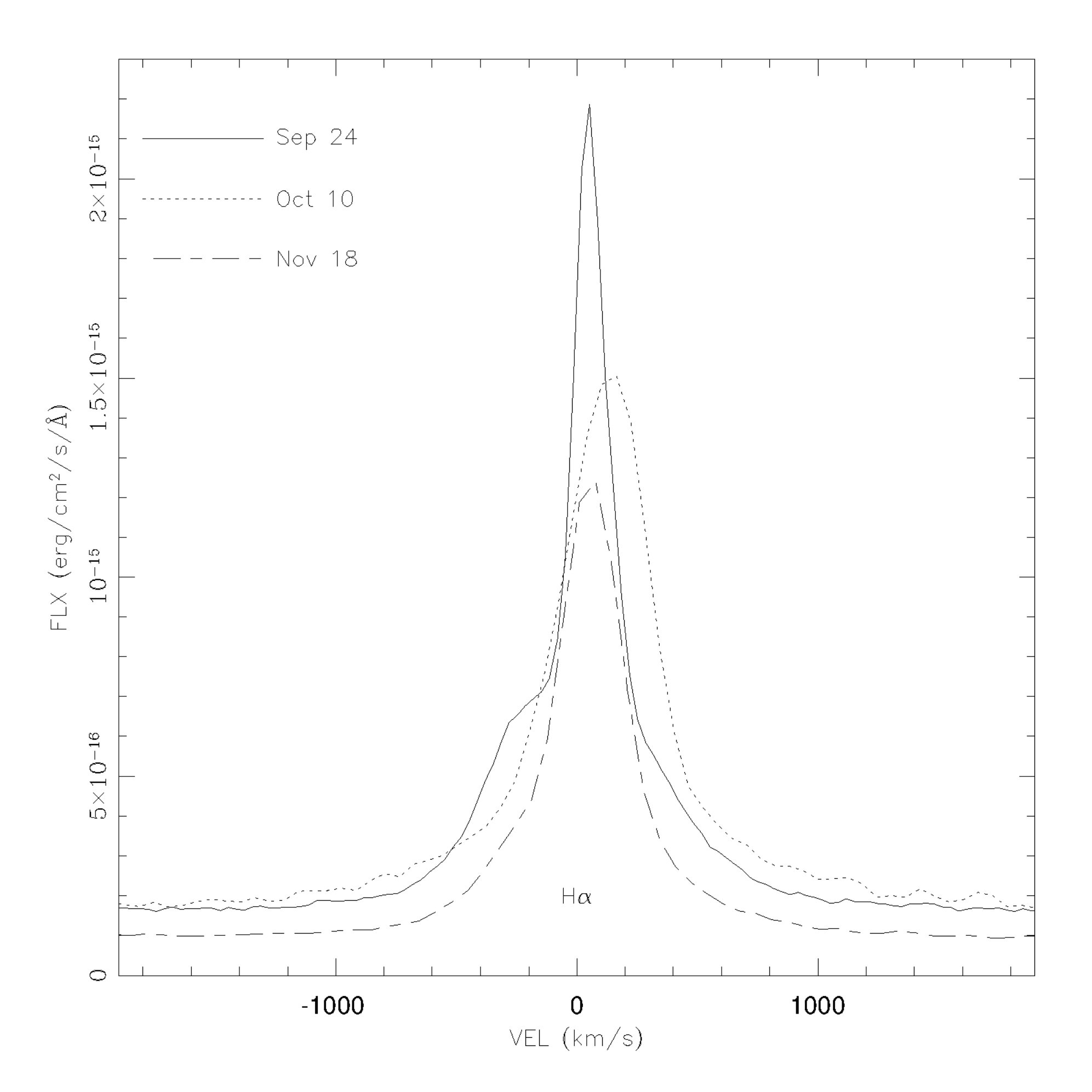}}\\
\resizebox{9.0cm}{!}{\includegraphics{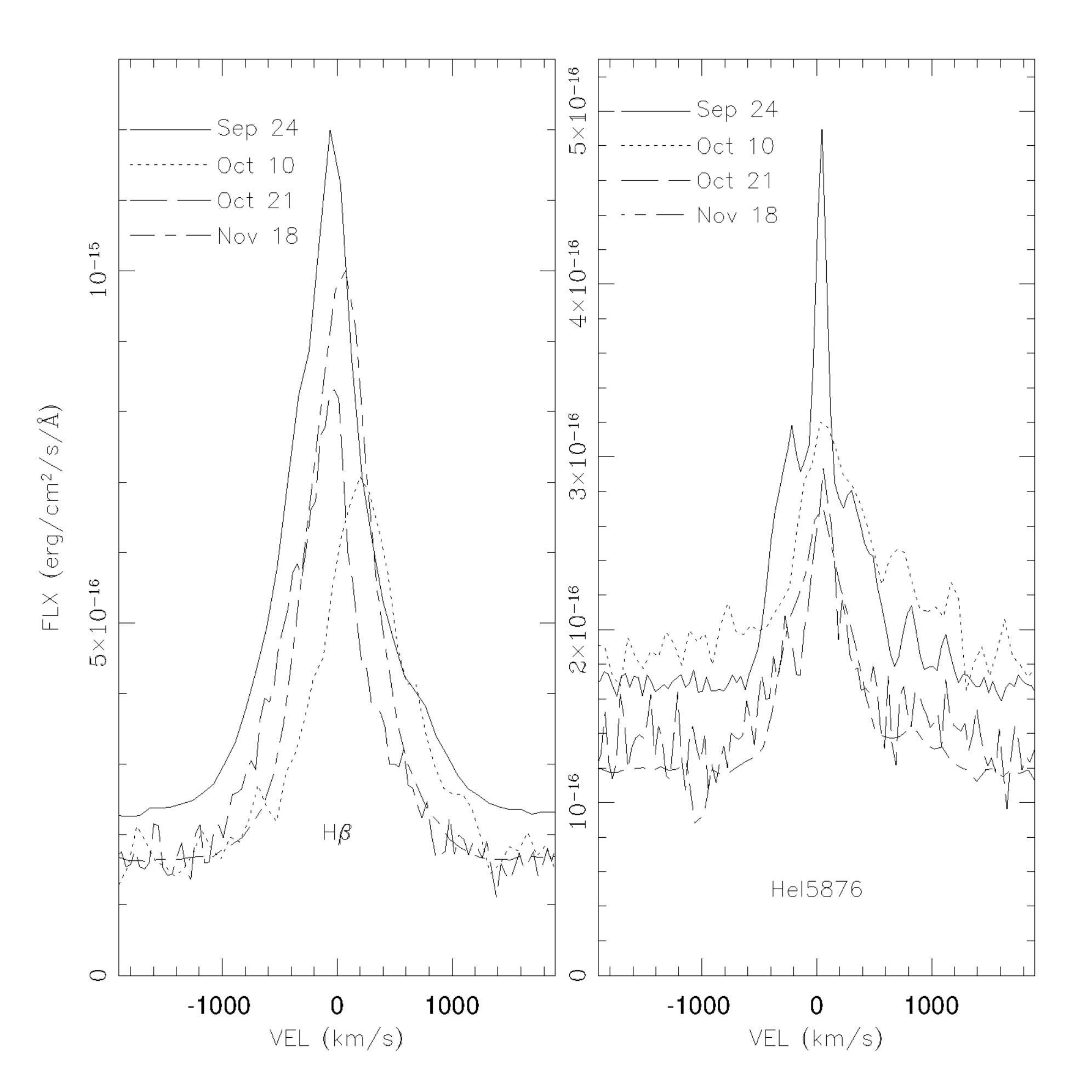}}
\caption{\label{linePROF} Line profile of H$\alpha$, H$\beta$, and HeI\,$\lambda$5876 at the various epochs. }
\end{figure}

\section{Analysis of the spectra}
\subsection{The single spectra}

Figure \ref{flxedSPC} shows that VY~Scl low state spectra are dominated by H and HeI emission lines over a slightly blue continuum that culminates in the Balmer jump in emission. By comparison of the line strengths and the 5000\AA~ continuum level of all the spectra, and of  the approximately constant low state V-band magnitude (V$\sim$18.1), we believe that all of our observations sample states of similar mass transfer rate (see Fig. \ref{flxedSPC}). The small flux differences are easily explained by the uncertainties in the calibration intrinsic to slit observations. In particular, a difference in the continuum slope of the 10 October spectrum seems compatible with colour losses due to the fixed east--west orientation of the slit and the timing of the science and the standard star observations. 

We compare the line profiles of the Balmer (H$\alpha$ and H$\beta$ and the HeI\,$\lambda$5876 emission lines  in Fig. \ref{linePROF}. The majority of the adopted instrument set-ups at the different instruments/telescopes have comparable resolution using grating with 600 lines/mm. The exceptions were the 10 October Goodman spectrum which used a grating with only 300 lines/mm (halving the resolution), and the 27 September FORS2 time series which adopted a grating with 
1200 lines/mm. The higher resolution of the latter was capable of  
separating well the HeI\,$\lambda$5876 and the Na\,I\,D lines. The figure 
shows that at all times the H and the HeI emission lines are characterised 
by broad wings extending up to $\sim\pm$1200 km/s from the line centre. 
Since these broad wings are consistent with accretion disc emission 
from the analysis of Doppler maps and trailed spectrograms 
(see Section 4), we must conclude that during this low state at least
some of the accretion disc is still present, perhaps with low-level ongoing
mass transfer as well.
Both the 10 October Goodman spectrum and the 27 September FORS2 time series extend 
to longer wavelengths, allowing us to detect spectral signatures of the 
secondary star, for example the redder slope and the TiO absorption 
band at $\sim$7150\AA. 
In neither case can we  properly fit a secondary star template 
spectrum (due to the low S/N and heavy fringing in one case or to 
the limited wavelength range in the other). However, the TiO band 
in the time series allows us to derive the radial velocity of the 
`back' of the secondary star thus constraining the mass ratio of 
the binary system (see Sec.\ref{sp}). 

The higher S/N ratio of the VLT spectra also allow us to identify with 
certainty  the weak emission line from MgI\,$\lambda$5184 \AA. 
The MgI\,$\lambda$5167 and 5172 are blended with the 
FeII\,$\lambda$5169 \AA. The FeII presence is suggested by the fact 
that the HeI lines at $\sim$4922 and 5015 \AA\, have somewhat broader 
full width at half maximum (FWHM) with respect to the HeI at 6678 \AA. In addition we detect the 
CaII~K line 3934 \AA\, in the spectrum from 24 September showing a smaller FWHM 
(about half that measured on the HeI lines). Following 
\citet{masonetal08-1}, we ascribe the Mg and Fe emission to the 
irradiated secondary, and CaII and Na\,I to stellar activity from a 
localised region in proximity of the L1 point. Mason et al.\ ascribed 
 the HeI emission to stellar activity as well. He I emission
likely consists of both secondary star and accretion disc/stream 
contributions since they show substructure and/or multiple components, 
and their FWHM values are not as narrow as those from MgI and CaII. 

\subsection{Time series spectroscopy}
In both time series spectroscopy runs,  all the emission lines
appear to be single-peaked while the Balmer and helium lines
show broad wings that are not evident in the metal lines.
The hydrogen and helium line profiles throughout the orbit are generally
complex from centre to wings and best fit by
multiple components. We fit them with one narrow
Gaussian and one or two additional components with Voigt profiles in
order to account for the extended wings.
The time-resolved spectral emission lines of the Na\,{\sc i}\,D doublet and
Fe\,{\sc ii}\,$\lambda$5169\footnote{This line is blending with MgI, as we said above. However, we will refer to this line as Fe\,{\sc ii} in the rest of the paper. The conclusions we draw are not affected by this.} show profiles that are very different in that they do not contain extended wings or a broad component.
Single Gaussians were used to fit these lines.

\subsubsection{Orbital period}
\begin{figure}
\resizebox{8.9cm}{!}{\includegraphics{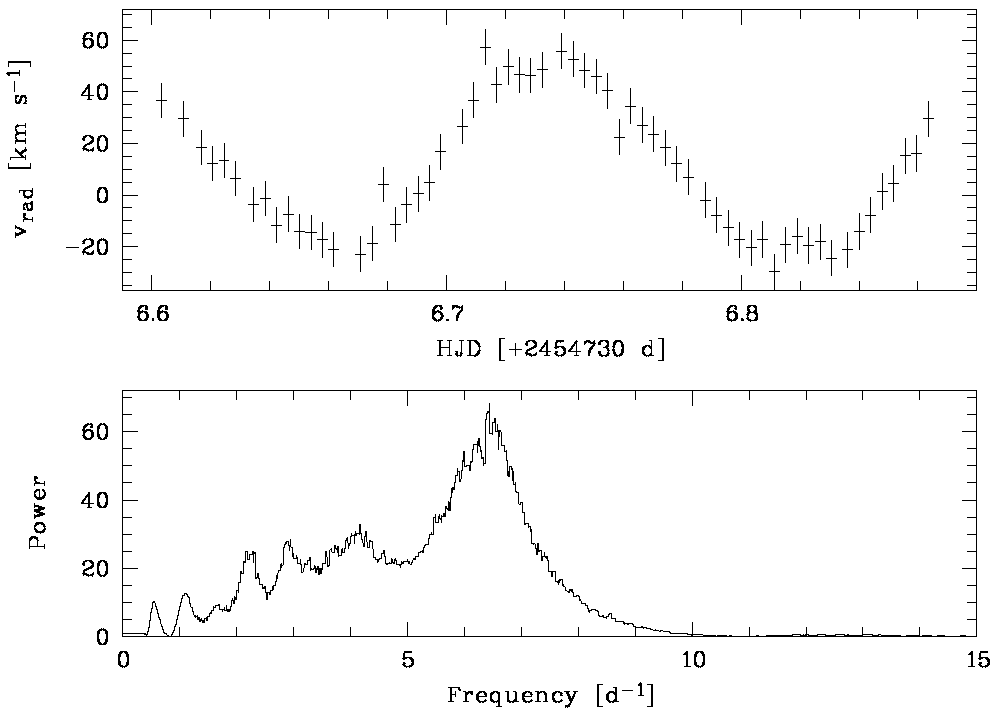}}
\caption{\label{Ha_sep} Top: Radial velocity curve of the
central H$\alpha$ emission line covering more than one orbit. Bottom:   Analysis-of-variance periodogram, the peak centred on 6.35\,d$^{-1}$. See text for details.}
\end{figure}

In an attempt to measure the radial velocities and to determine
the orbital
period, we started with the narrow line components of H and He and
fit a Gaussian function to them.
These radial velocity measurements show that only the FORS2 run covers more than one orbit
and that only the H$\alpha$ emission line, thanks to its higher signal,
shows a clear and smooth sinusoidal curve which is suitable for unambiguous
period analysis (Fig.\,\ref{Ha_sep}, top panel).
We used the AOV algorithm
\citep{schwarzenberg-czerny89-1} as implemented in MIDAS to create a
periodogram (see bottom panel of Fig. \ref{Ha_sep})
for H$\alpha$ and, using a Gaussian function, obtained the best fit period of $P_{\rm orb} = 3.8 \pm 0.3$\,h, where the uncertainty
corresponds to the $\sigma$ of the Gaussian fit. We note that the
periodograms for the other emission lines in the FORS2 data set
(i.e. He\,{\sc i} 5876, 6678, 7065, and Na\,{\sc i}\,D) are  noisier, but they deliver
consistent results within their respective uncertainties. 
We attempted to combine the FORS2 and FORS1 data to increase the accuracy of the orbital period; however, due to the large time difference between the two epochs, we obtained a number of narrow alias peaks all overlapping the area of the broad peak in Fig. \ref{Ha_sep}, which provided no additional constraints.

Past period determinations are those of \cite{hutchings+cowley84-1} and \cite{martinez-paisetal00-1};  the former  is based on intermediate state spectra (VY\,Scl was at about V$\sim$16\,mag) and the latter on data taken during a high state.
The two groups find significantly different orbital periods ($\sim$4.0 and $\sim$5.6\,h, respectively), but the best period determined by  \cite{hutchings+cowley84-1} is in relatively good agreement with our determination, given the uncertainties. However, we should note that neither the \cite{hutchings+cowley84-1} nor the  \cite{martinez-paisetal00-1} data sets covered a full binary orbit within one night\footnote{More precisely, \cite{hutchings+cowley84-1} do not present a log of the observations; they observed during two nights and mention that their data base is not long enough to determine unambiguously an orbital period.}, but combined data from different/consecutive nights. This introduces aliases which can possibly hide the true period.
Since our FORS2 data set covers more than a complete orbit continuously, we can plainly
rule out the value of $P=5.6$\,h by \cite{martinez-paisetal00-1}.

Looking at the data of Mart\'\i nez-Pais et al.\  in more detail
and  at the periodogram that they present in the paper, we notice several alias peaks, each of which can correspond to the orbital period. Only one of these peaks,  the one at 3.84(3)\,h, is consistent with our data. Thus, combining our
period which is robust but has a large uncertainty with the better defined but less robust period determination of Martinez-Pais et al., we conclude that the orbital period of
VY\,Scl is 3.84(3)\,h (0.160\,d)
and adopt this value for all  further analyses.

\subsubsection{Radial velocity curves}
\begin{figure}
\resizebox{8.9cm}{!}{\includegraphics{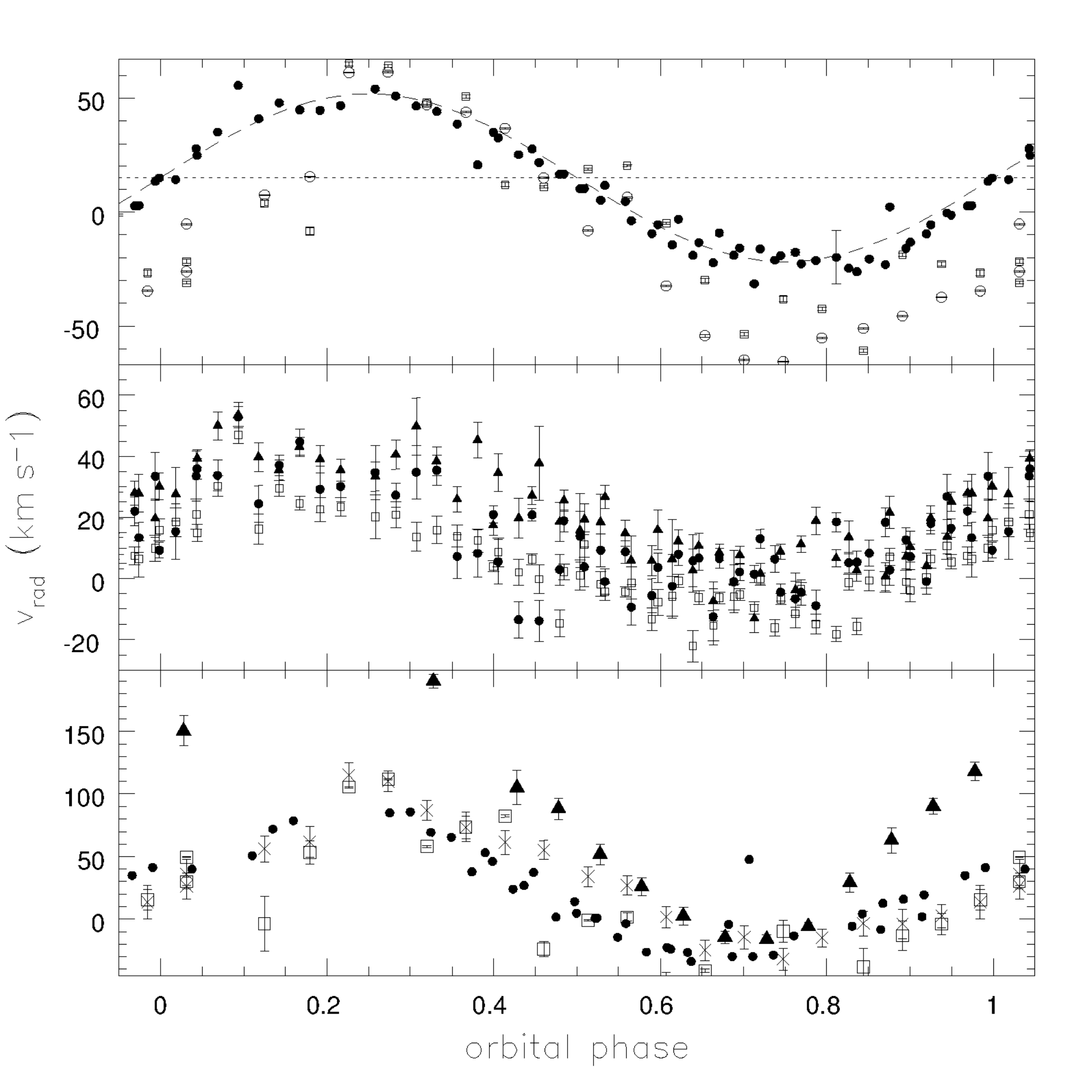}}
\caption{\label{rv_phase} Radial velocity curve of the observed emission lines and the TiO absorption. Top panel: Balmer lines: FORS2 H$\alpha$ measurements are marked by black filled circles, FORS1 H$\alpha$ measurements by empty circles, and FORS1 H$\beta$ measurements by empty squares. Middle panel: Radial velocity of the three HeI emission lines as measured in the FORS2 spectra. Bottom panel: Radial velocity measurements of TiO bandhead (black triangles), NaI D emission (black circles), FeII 5169 blending with MgI (crosses), and MgI 5183 (empty squares).}

\end{figure}

\begin{table}
\footnotesize
 \centering
  \caption{\label{fit_par} Best fit parameters for the radial velocity curves
for each line at each epoch. T$_{R/B}$ + -2454730 days gives the HJD of the zero phase, defined
as the blue-to-red
crossing of the central H$\alpha$ line, while $\Delta\phi$ corresponds to the phase offset with respect to
this zero phase.
}
  \begin{tabular}{@{}lllllc@{}}
  \hline
 epoch & line ID & K & $\gamma$ &  T$_{R/B}$ & $\Delta\phi$ \\
& & (km/s) & (km/s) & & \\
\hline
27/09/2008 & H$\alpha$ & 36.8$\pm$0.6 & 15.0$\pm$0.4 & 6.6983 & 0.0 \\
27/09/2008 & He\,{\sc i} 5875 & 18.0$\pm$0.7 & 20.9$\pm$0.5 &  & -0.07 \\
27/09/2008 & He\,{\sc i} 6678 & 17.6$\pm$1.0 & 15.1$\pm$0.7 &  & -0.09 \\
27/09/2008 & He\,{\sc i} 7065 & 20.3$\pm$0.9 & 23.6$\pm$0.6 &  & -0.04\\
27/09/2008 & Na\,{\sc i} 5890/6 & 50.9$\pm$2.3 & 14.6$\pm$1.5 &  & -0.05 \\
27/09/2008 & TiO &114$\pm$20 & na $\pm$13 & & -0.05 \\
18/11/2008 & H$\alpha$ & 56.5$\pm$1.7 & -7.0$\pm$1.2 & 58.6494 & 0.0 \\
18/11/2008 & H$\beta$ & 45.5$\pm$3.1 & -4.7$\pm$2.2 &  & 0.03 \\
18/11/2008 & Fe\,{\sc ii} 5169 & 57.2$\pm$2.3 & -43.6$\pm$1.6 &  & -0.03 \\
\hline
\end{tabular}
\end{table}

The radial velocities of all emission lines were phased using
the 3.84\,h period determined above and were fit (Fig. \ref{rv_phase})
with the function $v_r(\phi) = \gamma + K \sin{2\pi \phi}$ using a
Monte Carlo technique to determine the uncertainties of the individual parameters. The parameters of the best fit for each line are
listed in Table\,\ref{fit_par}. We define the epoch of zero-phase
as the best fit blue-to-red crossing of the narrow central
H$\alpha$ component. This corresponds to $\rm HJD_0 = 2454736.6983(4)\,d$
and $\rm HJD_0 = 2454788.6494(7)\,d$ for the FORS2 and FORS1 run, respectively.
The phase offsets in Table\,\ref{fit_par} have been computed with
respect to these ephemerides.

\begin{figure*}
\centerline{
\resizebox{!}{8.3cm}{\includegraphics{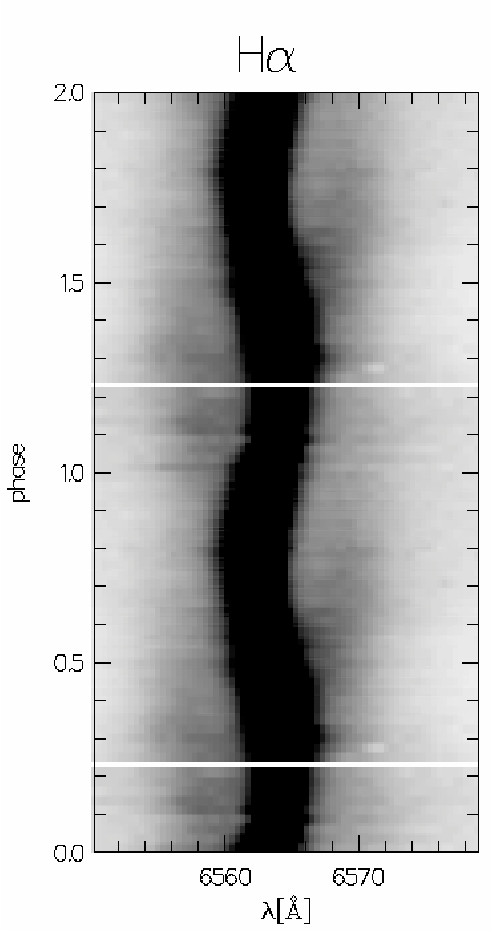}} ~
\resizebox{!}{8.3cm}{\includegraphics{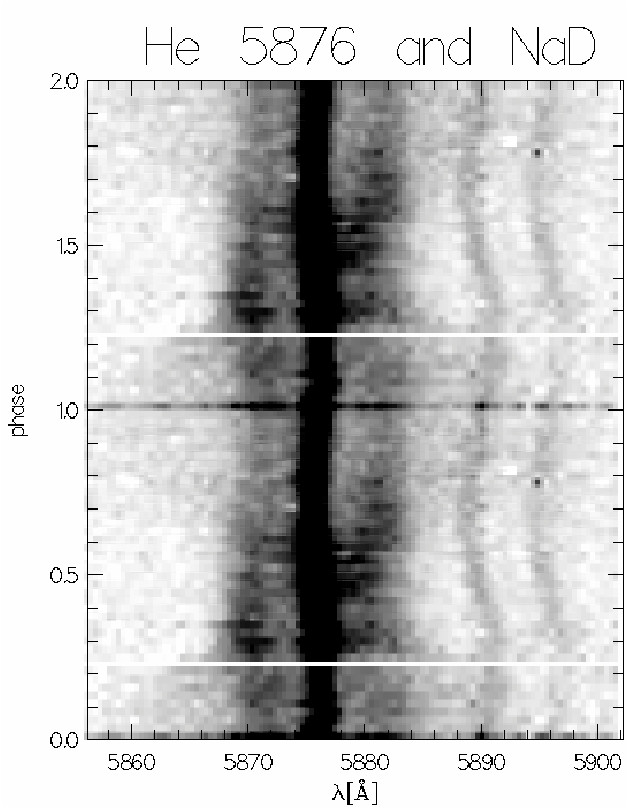}} ~
\resizebox{!}{8.3cm}{\includegraphics{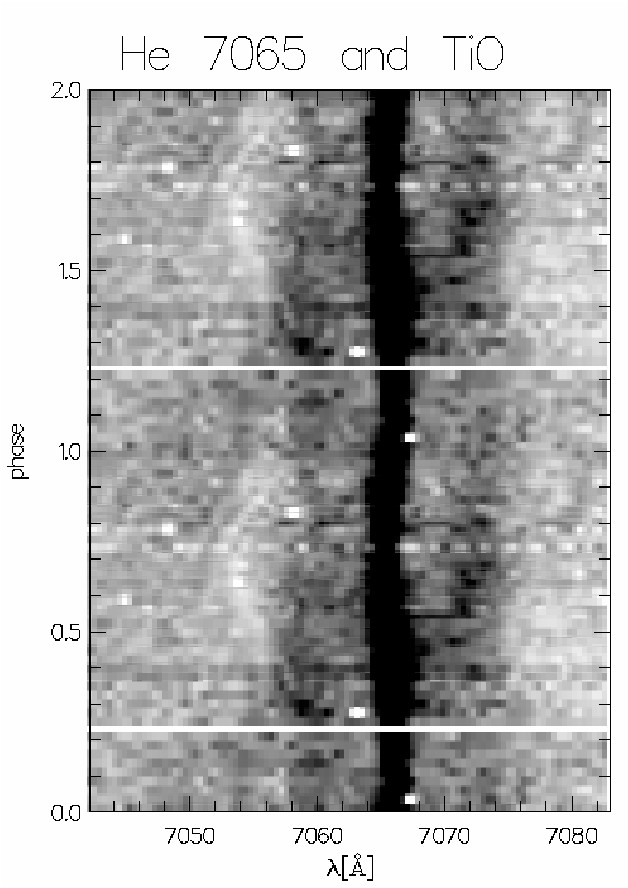}}}
\caption{\label{trailed} Trailed spectrograms of the FORS2 data showing
the regions around H$\alpha$, He{\sc i} with Na\,I\,D, and He{\sc i} with
TiO head. The greyscale is inverted, i.e. dark shades of grey represent
emission and the TiO absorption appears lighter than the continuum.
For clarity, a whole orbital cycle is repeated. }
\end{figure*}

\begin{figure}
\centerline{
\resizebox{!}{8.7cm}{\includegraphics{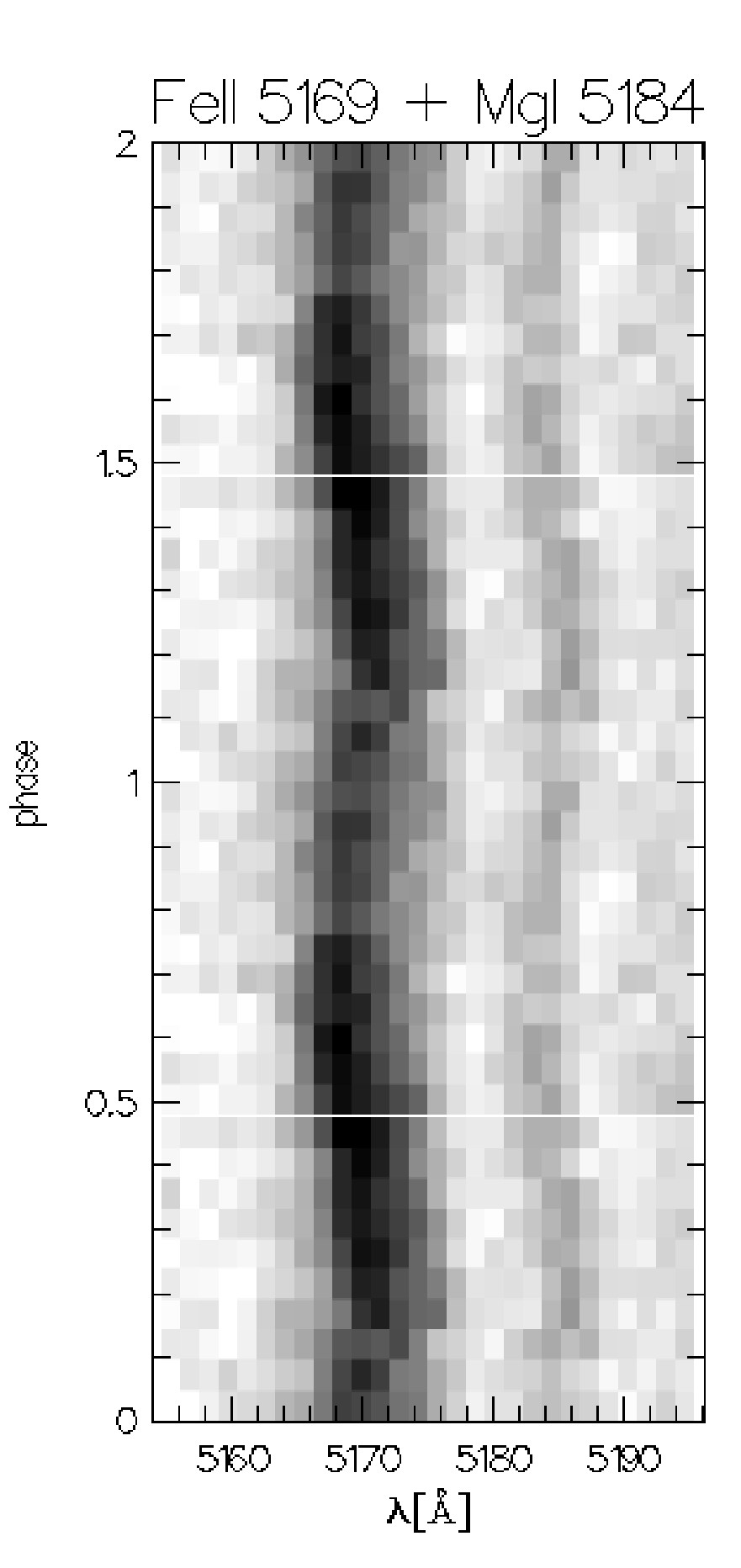}} ~
\resizebox{!}{8.7cm}{\includegraphics{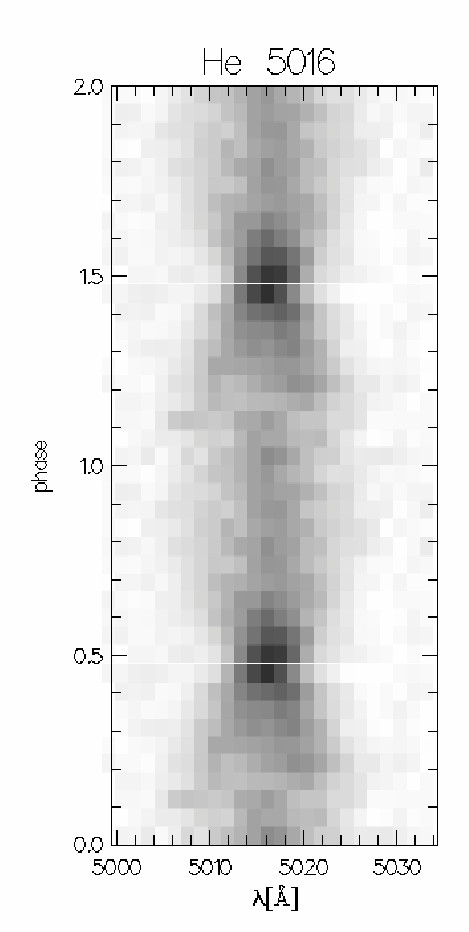}}} ~
\caption{\label{trailed_elena} Trailed spectrograms of the FORS1 data
showing the different behaviour of the Fe{\sc ii} and He{\sc i} lines. The weak emission on the red side of the 5169 \AA\ emission is MgI 5184 \AA}
\end{figure}

\begin{figure}
\resizebox{9.0cm}{!}{\includegraphics{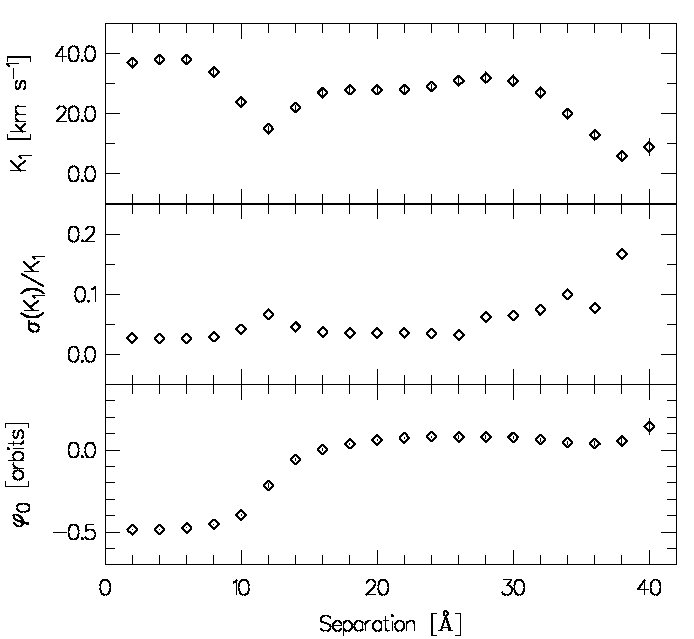}}
\caption{\label{diag} H$\alpha$ diagnostic diagram:  radial velocity
amplitude $K$, its uncertainty, and
the offset of the orbital phase (red to blue crossing yields $\phi_0 =0$) 
plotted against the separation of the two Gaussians used for the fitting
(see text for details). For small separations, the Gaussians trace the central peak of the emission line and the resulting parameters are close to those
listed in Table \ref{fit_par}. For separations between 16 and 26\,\AA , the
fitting gives stable parameters while the wings are traced. Further outside,
the noise increases and the parameters start varying accordingly.
}
\end{figure}

We note that we did not attempt to fit the radial velocities of the He\,{\sc i} lines in
the FORS1 data set because at lower resolution the central peaks
are severely affected by the broad component/wings. We did, however, fit
the weak TiO bandhead at $\sim$7054\,\AA \ which we detected on the
trailed spectrogram of the He\,{\sc i}\,$\lambda$7065 line
(see Fig.\ \ref{trailed}, right side) and the wings of the H$\alpha$ emission line in the higher resolution FORS2 spectra.

As the TiO absorption cannot be easily spotted in each single spectrum,
we measure its radial velocity from the cursor position in the wavelength
calibrated trailed image of the spectra. We independently determined the
radial velocity in each spectrum seven times and averaged these values together. We note that
the TiO absorption was not visible at all orbital phases because of the `disturbing' emission from the He\,{\sc i}\,$\lambda$7065 extended
wings.
The radial velocity curve of the TiO bandhead is plotted
in Fig.\ \ref{rv_phase} and the parameters for the best fit are
listed in Table \ref{fit_par}.
As no precise zero-wavelength could be attributed to the TiO headband, we do not derive a 
system velocity from the radial velocities of this spectral band and only 
give the uncertainty in the shift to judge the quality of the fit. 
In Fig.\ \ref{trailed} the trailed spectrogram of the TiO absorption 
bandhead is plotted in comparison with the various 
emission lines in the same FORS2 time series. 

The radial velocities of the H$\alpha$ line wings were measured using the
double-Gaussian fit method described in \citet{schneider+young80-2} and
\citet{shafter83-1}.
Two Gaussians, each  4\,\AA\ in width, were fit to the line profile;
the separation of the two Gaussians was varied between 2 and 40\,\AA.
In Fig.\ \ref{diag}, the parameters $K$ and $\phi_0$ are plotted against
the separation of the two Gaussians.
We note that the phase offset $\phi_0$ in the diagnostic diagram
refers to zero phase being the red-to-blue crossing of the line as the
diagram is computed to find a radial velocity fit for the primary star.
Therefore, $\phi_0 = -0.5$ indicates that the lines are crossing from
blue to red at zero phase $\phi = 0$.
For small separations, the two
Gaussians trace the central peak of the line, and the results
of the radial velocity fits are similar to those derived for the
radial velocities of the central line that are given in Table \ref{fit_par}.
For separations between 16 and 26\AA , the fitting of the radial velocities
gives stable and robust parameters while tracing the outer line wings,  values that are a proxy of the white dwarf motion.
Averaging over this range, we derive the radial velocity amplitude
$K_1 = 29 \pm 2 \rm ~ km\,s^{-1}$ and a phase offset of $\phi_0 = 0.06 \pm 
0.02$.

Similar measurements of the He\,{\sc i} line wings in the FORS2 data
set did not result in robust fit parameters
for any Gaussian separation. This is probably due to a combination of the
lower S/N ratio and the stronger variation of the  He\,{\sc i}
emission line wings.  As seen from the trailed spectra in Figs.\ \ref{trailed} and \ref{trailed_elena}, the He lines are of variable extension and profile across
the orbit. For any separation below $\approx 20$\,\AA\
the double Gaussian thus traces different emitting regions at
different orbital phases.  Further outside, where the line is bound to become
more stable, the S/N  becomes too low for reliable measurements.

\subsection{Characterising the distribution of the emission line sources}

The analysis of the radial velocity curves and their best fits show that
the narrow emission component of H$\alpha$, H$\beta$,
He\,{\sc i}\,$\lambda$5876, 6678, and 7065; the
Fe\,{\sc ii}\,$\lambda$5169 and  Na\,{\sc i}\,D emission
lines; and the TiO bandhead absorption all move in phase.
As the TiO absorption
can clearly be attributed to the secondary star, all of the
narrow emission line components also have a likely origin somewhere in the secondary star's regime. 
The radial velocity amplitude $K$ is different for all these lines, which implies that they originate from different line emitting regions. The velocity amplitudes follow the series
$K_{\rm HeI} < K_{\rm H\alpha} < K_{\rm NaI} < K_{\rm FeII} < K_{\rm TiO}$, 
a sequence one would expect for a co-rotating, irradiated secondary star with a temperature gradient on its surface. The hottest part of the secondary is the L1 region. This is where He\,{\sc i} is emitted, which thus has the smallest 
radial velocity amplitude. The H$\alpha$ and H$\beta$ line emitting regions will be larger, extending up to colder regions and higher velocities. The coldest
region of the secondary star is the non-illuminated backside, where the TiO bands form. Their velocity amplitude is thus expected to have the highest value, which is indeed what we observe.

Temperature gradients in line emitting regions of the secondary
are not uncommon among CVs and have been observed,
for example in BB\,Dor during a low state \citep{schmidtobreicketal12-1}.
The BB\,Dor H$\alpha$, He\,{\sc i,} and Na\,{\sc i}\,D narrow emission lines 
and TiO aborption bands follow a sequence similar to those observed here, and
were explained using an irradiation-introduced temperature
gradient on the secondary star.

The Na\,{\sc i} emission, and possibly the Fe\,{\sc ii} or Mg\,{\sc i}, are
associated with chromospheric activity.
They are thus expected to form more or less uniformly  all over
the active secondary star, and therefore represent the best tracers
of the secondary star's orbital motion \citep[see][]{howelletal00-1,masonetal08-1} 

\begin{figure}
\centerline{
\resizebox{4.4cm}{!}{\includegraphics{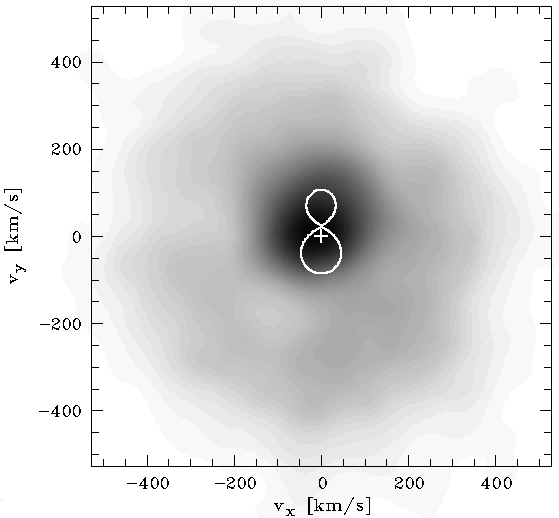}}
\resizebox{4.4cm}{!}{\includegraphics{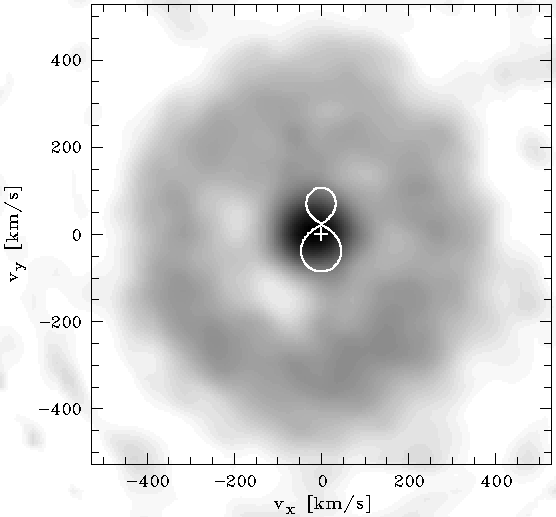}}
}
\caption{\label{doppler} Doppler maps of VY\,Scl showing the distribution of
H$\alpha$ emission (left plot) and the averaged He\,I emission (right plot)
in velocity coordinates. The white cross marks the centre of
rotation. The maps are oriented  such  that the phase angle
is zero towards the top and increases clockwise.}
\end{figure}

From the diagnostic diagram of the H$\alpha$ line (and the discussion above), we determine that the line wings form in a region which is moving exactly
anti-phased with respect to the one giving rise to the
narrow emission lines. 
The line wings therefore originate in
an approximately symmetric structure on the side of the primary star, likely an accretion
disc.
In order to better locate this line forming
region we apply the Doppler tomography method using the code of \citet{spru98}
with a {\sc MIDAS} interface replacing
the original {\sc IDL} routines \citep{tappertetal03-1}.
Making use of the rotation of the binary system, the calculated
Doppler maps $I(v_x,v_y)$ display the intensity of the line component
emitted at a velocity $(v_x,v_y)$.
In Fig.\ \ref{doppler}, the Doppler maps of the H$\alpha$ and He\,{\sc i}
lines are plotted.
Both maps similarly show two features: a strong peak just above the centre
of mass and an almost homogeneous ringlike structure.
The strong peak is associated with the narrow line components that show the
low radial velocity amplitude of the L1 and close-by regions.
The ringlike structure in velocity space is the sign of emission
from an accretion disc.
This is in agreement with the findings of \cite{hamilton+sion08-1} who
also observed VY\,Scl during a low state (V-magnitude was fainter than 15.8 mag), and
found a significant inner accretion disc contribution by modelling the IUE UV spectra.

We point out that the observation of a disc (or at least a small inner disc) is an argument against the
`no disc option' proposed by \citet{hameury+lasota02-1}.
Hamureuy and Lasota estimate that the value of the magnetic moment necessary
to prevent the formation of a disc 
is in the range between DQ\,Her and intermediate polars, i.e. $B  > 6$\,MG. Thus, with the observed presence of at least a partial accretion disc, we can
rule out a high magnetic field for VY\,Scl.

\subsection{System parameters}\label{sp}
A determination of the centre position of each Doppler map yields the velocity amplitude of
$K_1 = 28 \pm 5\rm ~km\,s^{-1}$, a value in good agreement
with the value $K_1 = 29 \pm 2\rm ~km\,s^{-1}$
from the diagnostic diagram. We thus adopt this latter value as the projected
velocity $v \sin{i}$ of the white dwarf. 

For the secondary star, different values for radial velocity amplitude
have been measured 
depending on the origin of the line emitting region.  The velocity 
amplitude derived from
the TiO bands $K_{\rm TiO} = 114\pm20\,\rm km\,s^{-1}$ represents a strict 
upper limit for the radial velocity amplitude $K_2$ of the secondary.
For a thorough discussion of the effect of irradiation on the velocity 
amplitude of absorption lines, see \cite{wade+horne88-1}. 

Assuming that the Na\,{\sc i} and Fe\,{\sc ii} or Mg\,{\sc i} emission lines
come from chromospheric activity uniformly distributed over the secondary star, 
they are the best tracers of the secondary star's orbital motion and we 
derive $K_2 = 54 \pm 3\rm ~km\,s^{-1}$ as their average value. However, 
one might argue that the Na\,D and Fe\,II lines
can be affected by irradiation and/or that the 
chromospheric emission sources are clustered towards the L1.
In that case, this value would only represent a lower value for
the secondary's radial velocity amplitude.

Approximating the geometry of the elongated, Roche-lobe filling star with 
an ellipse, we can easily
calculate the distance between the secondary's centre of mass and the binary's 
centre of mass as the average of the 
L1 distance and its backside distance. 
Since the velocity on a co-rotating structure increases linearly with the 
distance, we can approximate the velocity at 
the centre of mass of the secondary as the average between the velocity 
of L1 and the velocity at the backside of the secondary. The radial 
velocity of L1 can be estimated as the velocity amplitude 
$K_{\rm HeI} = 19\,\rm km\,s^{-1}$ of the He\,{\sc I} line, that of the 
backside from the $K_{\rm TiO} = 114\pm20\,\rm km\,s^{-1}$. For the 
secondary's radial velocity, this approximation yields $K_2 = 67\pm11\,\rm km\,s^{-1}$.
This value is slightly higher than that derived from the chromospheric lines, indicating that the line emitting regions are indeed shifted towards the L1.

Using this geometrically derived radial velocity as the most likely value for the 
secondary's orbital motion, we calculate the mass ratio of the binary
as $M_2/M_1 = K_1/K_2 = 0.43 \pm 0.15$.
With an estimated secondary star mass of 
$M_2 = 0.3-0.35$\,M$_\odot$ \citep[eq.6]{howelletal01-1}, we 
derive  the white dwarf mass to $M_1 = 0.5-1.1$\,M$_\odot$ and an inclination around $15^\circ$.

\section{Summary and conclusions}
We observed  VY\,Scl in optical spectroscopy on several epochs
during its 2008 low state. At all times, it appears that at least a small inner disc was present.
This is supported by the H\,$\alpha$ and
He\,{\sc i} Doppler maps that show the disc-like structure around the
position of the white dwarf. The presence of such an inner accretion disc in
the low state rules out the possibility of VY\,Scl containing a highly magnetic white dwarf and being
an intermediate polar.
The TiO absorption bands, the narrow central part of the Balmer and He\,I emission lines, and the 
emission lines in Na\,I and Fe\,II all move in phase and are anti-phased to the broad H\,$\alpha$ and
He\,{\sc i} line wings that result from the accretion disc. Different radial velocity amplitudes of these lines trace the temperature profile on the irradiated secondary.

The time-resolved spectroscopic observations together with old data from \cite{martinez-paisetal00-1}
allowed us to unambiguously determine the orbital
period of the binary: $P=3.84\pm0.03$\,h. 
We used the broad line wings to trace the white dwarf movement and
derived  $K_1 = 29 \pm 2\, \rm km s^{-1}$. 
The secondary star
shows a temperature gradient as expected from irradiation. 
From geometric considerations, we derive the most likely value 
$K_2 = 67 \pm 11\, \rm km s^{-1}$ and $M_2/M_1 = 0.43$.
With the secondary's mass between 0.3 and 0.35\,M$_\odot$, as expected for a 
Roche-lobe filling secondary in a $P=3.84$\,h-system, we calculate 
the mass of the white dwarf $M_1= 0.5-1.1$\,M$_\odot$.
With about $15^\circ$, the inclination of VY\,Scl is very low,
which makes it difficult to gain any more detailed information from the
radial velocities.

\section*{Acknowledgements}
We are very grateful to Pablo Rodriguez, Tom Marsh, Antonio Bianchini, Claus Tappert, and Bob Williams who provided valuable input during long discussions.
The use of {\tt MOLLY} developed
by Tom Marsh is gratefully acknowledged.
We acknowledge with thanks the variable star observations from the AAVSO International Database contributed by observers worldwide and used in this research.
The SOAR telescope is  funded by a partnership of the Minist\'{e}rio
da Ci\^{e}ncia, Tecnologia, e Inova\c{c}\~{a}o (MCTI) da Rep\'ublica
Federativa do Brasil, the U.S. National Optical Astronomy Observatory,
the University of North Carolina at Chapel Hill, and Michigan State University.
Access to SMARTS has been made possible by support from the Provost and the
  Vice President for Research of Stony Brook University.
The research leading to these results has received funding from the European Research Council under the European Union’s Seventh Framework Programme (FP/2007– 2013) / ERC Grant Agreement n. 320964 (WDTracer).
LS thanks the STScI for the hospitality during the scientific stay which
made this analysis possible, and acknowledges the support through the
ESO DGDF programme. EM thanks ESO for the hospitality and the support in May 2017, when this work was finalised

\bibliographystyle{aa} 
\bibliography{aamnem99,aabib}
\end{document}